\documentstyle[11pt,aaspp4]{article}
\slugcomment{Submitted to {\it The Astronomical Journal}}
\lefthead{Philp et al.\ }
\righthead{Do OB Runaway Stars Have Pulsar Companions?}

\begin{document}

\title{Do OB Runaway Stars Have Pulsar Companions?}
\author{Colin J. Philp \altaffilmark{1} and Charles R. Evans \altaffilmark{2}}
\affil{Department of Physics and Astronomy \\
 University of North Carolina, Chapel Hill, NC 27599-3255}
\altaffiltext{1}{\footnotesize E-mail: tas@physics.unc.edu}
\altaffiltext{2}{\footnotesize E-mail: evans@physics.unc.edu}
\author{Peter J. T. Leonard \altaffilmark{3}} \affil{Department of
Astronomy \\ University of Maryland, College Park, MD 20742}
\altaffiltext{3}{\footnotesize E-mail: pjtl@astro.umd.edu}
\author{Dale A. Frail \altaffilmark{4}} \affil{National Radio
Astronomy Observatory, \altaffilmark{5} Socorro, NM 87801}
\altaffiltext{4}{\footnotesize E-mail: dfrail@nrao.edu}
\altaffiltext{5}{\footnotesize Operated by Associated Universities,
Inc.\ under cooperative agreement with the National Science
Foundation}

\keywords{Pulsars: General -- Stars: Early-Type -- Stars: Runaway Stars}

\begin{abstract}

We have conducted a VLA search for radio pulsars at the positions of
44 nearby OB runaway stars.  The observations involved both searching
images for point sources of continuum emission and a time series
analysis.  Our mean flux sensitivity to pulsars slower than 50 ms was
0.2 mJy.  No new pulsars were found in the survey.  The size of the
survey, combined with the high sensitivity of the observations, sets a
significant constraint on the probability, $f_p$, of a runaway OB star
having an observable pulsar companion.  We find $f_p \le 6.5$\% with
95\% confidence, if the general pulsar luminosity function is
applicable to OB star pulsar companions.  If a pulsar beaming fraction
of \onethird\ is assumed, then we estimate that fewer than 20\% of
runaway OB stars have neutron star companions, unless pulsed radio
emission is frequently obscured by the OB stellar wind.  Our result is
consistent with the dynamical (or cluster) ejection model for the
formation of OB runaways.  The supernova ejection model is not ruled
out, but is constrained by these observations to allow only a small
binary survival fraction, which may be accommodated if neutron stars
acquire significant natal kicks.  According to Leonard, Hills and
Dewey (1994), a 20\% survival fraction corresponds to a 3-d kick
velocity of 420 km s$^{-1}$.  This value is in close agreement with
recent revisions of the pulsar velocity distribution.

\end{abstract}

\section{Introduction}

     The OB runaway stars are massive young stars that have high
peculiar velocities ($|V_p| > 30$ km s$^{-1}$) and/or large scale
heights above the Galactic plane (e.g., $> 100$ pc).  Such stars stand
out since ordinary OB stars exhibit a scale height and velocity
dispersion more typical of interstellar gas ($\simeq 50$ pc and
$\simeq 10$ km s$^{-1}$ respectively; Mihalas \& Binney 1981).  Their
name derives from the fact that some of them appear to be ``running
away'' from massive star forming regions (Blaauw \& Morgan 1954;
Blaauw 1989).

The two most frequently invoked explanations for the origin of OB
runaways are the supernova ejection model (Blaauw 1961; Stone 1991)
and the cluster, or dynamical, ejection model (Poveda, Ruiz \& Allen
1967; Gies \& Bolton 1986).  In the first scenario, the more evolved
of two OB stars in a binary system undergoes a supernova explosion
which imparts a runaway velocity to its companion due to momentum
conservation.  Although the initially more massive primary star will
be the first to explode, post main-sequence mass loss will tend to
reverse the mass ratio and circularize the orbit.  Thus, it is likely
that the newly formed neutron star (NS) will remain bound to the
surviving OB star, since half the system mass must be lost to unbind a
circular binary.  Even a ``natal'' kick of 100 km s$^{-1}$ given to
the NS (due to an asymmetric supernova) will leave most OB runaways
with bound NS companions (Leonard \& Dewey 1993).  The newly-formed
binary will acquire a significant orbital eccentricity,
$\epsilon^{'}$, and runaway velocity, $V^{'} = \epsilon^{'} \times
V_{1,{orb}}$, where $V_{1,{orb}}$ is the pre-supernova orbital
velocity of the primary (Dewey \& Cordes 1987).  In contrast,
dynamical ejection relies on close dynamical interactions
(scatterings) involving OB binaries in young open clusters.  The
resulting three- or four-body interactions can eject stars at runaway
velocities.  This mechanism predicts that fewer than $10\%$ of OB
runaways will have companions of any kind, compact or otherwise
(Leonard \& Duncan 1988, 1990).

Studies aimed at trying to determine the origin of the OB runaways
have relied on the binary fraction to distinguish between the above
two mechanisms.  Gies and Bolton (1986) conducted a search for time
variations, indicative of orbital motion about a compact companion, in
the radial velocities of 36 bright OB runaways.  They also used
existing X-ray observations of their candidates to search for evidence
of accretion onto a NS or black hole.  They found no evidence for
compact companions which led them to favor dynamical ejection.
Leonard \& Dewey (1993) used a Monte Carlo program to simulate OB
runaway production by the supernova ejection mechanism.  They found
that the upper envelope of runaway velocity is anti-correlated with
mass of the progenitor system, fewer than 10\% of O stars will have
peculiar velocities greater than 50 km s$^{-1}$, and the majority of
runaways will have NS companions.  The authors conclude that the
observations of OB runaways are better explained by dynamical ejection
than supernova ejection.

Despite the above evidence for dynamical ejection, the recent
discoveries of PSRs B1259-63 (Johnston et al.\ 1992) and J0045-73
(Bell et al.\ 1995), both companions to B-type main sequence stars,
indicate that supernova ejection and NS retention do occur.  These
systems share much in common with the Be/X-ray binaries, a subgroup of
the high mass X-ray binaries (HMXB) (Bhattacharya and van den Heuvel
1991), whose orbits have long periods and are highly eccentric.
Indeed, PSRs B1259-63 and J0045-73 may be examples of the evolutionary
step that just proceeds the formation of Be/X-ray systems.  Additional
evidence for supernova ejection comes from the kinematics of the HMXB
(Brandt \& Podsiadlowski 1995) and from the existence of the binary
pulsars 1913+16 and 1534+12 which are thought to be an endpoint of
massive binary evolution.

If supernova ejection occurs and some neutron stars remain bound to OB
runaways, then a fraction, $f_p$, of the OB runaways will have pulsar
companions.  We conducted a search for pulsed and unpulsed radio
emission at the positions of 44 OB runaways.  Our search is sensitive
to pulsars in long period, eccentric orbits, far from the potential
obscuring effects of the OB stellar wind.  The Gies and Bolton search
would very likely have missed such objects because of the very small
variation in the OB star's velocity except for brief periods near
periastron.  Binary pulsars such as these may be common, but were
missed by previous pulsar surveys because of selection effects (for
example, PSR B1259-63 lies far out from the Galactic plane, has a
relatively short pulse period, has a binary companion, and has not
been detected at 400 MHz).  Our candidate selection criteria and
observations are discussed in section 2.  In section 3, we discuss our
sensitivity and results.  Section 4 contains a Bayesian statistical
analysis of our result and section 5 is a discussion of implications
for runaway star formation and neutron star natal kicks.

\section{Observations and Analysis}

Detecting a pulsar in the presence of a massive stellar wind
represents a two-fold problem.  Firstly, the ionized wind can both
scatter and absorb the radio beam coming from a pulsar, temporally
broadening the pulses, reducing the pulsar's luminosity, or possibly
eclipsing it completely.  Secondly, the wind may be a confusing source
of thermal or non-thermal (shock) radio emission (Bieging et al.\
1989).  In an attempt to circumvent these problems, we adopted a
two-part approach: imaging the field containing the OB star while
simultaneously collecting a time series.  These two aspects of the
search are complimentary. If pulses are smeared out by scattering in
the stellar wind, a pulsar may still be visible as a point source in
the continuum map.  We made images at both 1.4 and 5.0 GHz,
anticipating being able to distinguish pulsars from wind emission by
their tendency to have steep, negative spectral indices.  We searched
for pulsed emission at 1.4 GHz, rather than at a lower frequency, in
order to minimize the effects of free-free absorption
($\propto\nu^{-2}$) and temporal scattering ($\propto\nu^{-4.4}$;
Manchester \& Taylor 1977).  Previous pulsar searches conducted at
lower radio frequencies might have systematically missed objects like
PSR B1259-63 which was detected at 1.4 GHz but not at 400 MHz
(Johnston et al.\ 1992).

\subsection{Candidate Selection}

Table 1 lists our OB runaway program stars along with their distances,
peculiar velocities and spectral types.  They were chosen from four
sources: Bekenstein and Bowers (1974), Blaauw (1993), Conlon et al.\
(1990) and Gies and Bolton (1986). Our main selection criteria were
that the OB runaway be near by and that the spectral type should be no
earlier than O5.  The first criterion seeks to insure that any
potential pulsar companions will be near or above our detection
sensitivity.  For this reason, most of our program stars have
distances less than 1 kpc.  The second criterion attempts to minimize
the adverse effects of the stellar wind.  Mass flux from OB stars is a
strong function of their mass and luminosity (Abbott 1982).
Restricting the survey to B and late O stars reduces the chance of
pulsar obscuration and wind emission.  In addition to these criteria,
we attempted to select OB runaways that have very high velocities and
ones for which a birthplace in a star-forming region has been
suggested (see Blaauw 1993).

\subsection{Observations}

Observations were made in February, 1994, using NRAO's Very Large
Array (VLA) near Socorro, NM.  At the time of our observations, the
VLA was in the hybrid D/A configuration, giving us a mix of very long
and very short baselines.  We observed two polarizations at 1.4 and
5.0 GHz with 50 MHz of bandwidth in each polarization.  At 1.4 GHz, we
used the VLA in the phased array mode, simultaneously collecting
continuum visibility data and a time series.  The analog sum signal
was fed into a filter bank where each 50 MHz band was subdivided into
14 channels, each 4 MHz wide, to allow dedispersion of the data
offline.  The time series sample length was 2.6 ms.  Each target was
observed for a total of 12 minutes, yielding time series with 2$^{18}$
samples.  In addition to the OB runaways, we observed four known
pulsars for testing and calibration purposes (see Table 2).

\subsection{Data Analysis}

We carried out a standard Astronomical Image Processing System (AIPS)
reduction of the visibility data.  Our time series data analysis
involved four major steps.  In the first step, the time series were
dedispersed for a range of dispersion measures (DM) from 0 to 895 pc
cm$^{-3}$ in steps of 15.  In the second step, each dedispersed time
series was Fourier transformed using an FFT algorithm, a power
spectrum was computed and the spectrum was flattened and normalized by
a running average. The third step was to choose the strongest pulsar
candidate in the power spectrum.  Using the fact that pulsars
typically have many harmonics, the first half of the power spectrum
was stretched and added to the original so that the power in a given
frequency bin, $n_b$, would add to the power in bin 2$n_b$.  This
``harmonic folding'' was performed four times for each time series so
that the power in the first harmonic adds to the 2nd, 4th, 8th and
16th harmonics.  A best frequency, $f_{\rm best} = f_{\rm max} /
n_{\rm fold}$, was chosen where $f_{\rm max}$ is the frequency with
the highest signal to noise ratio (SNR) in any of the folded spectra
and $n_{\rm fold}$ is the folding number (i.e., 1, 2, 4, 8 or 16).  In
the final step, the dedispersed time series was divided into segments
of length $f_{\rm best}^{-1}$ and the segments were summed to produce
an integrated time profile.  Any profile having a pulsar-like shape
was examined as a possible pulsar candidate.

The VLA suffers as a pulsar detection instrument in that, being an
interferometer with 27 separate antennas, spurious periodic signals
are present in the data from many different electronic sources.
Consequently, narrow band interference must be removed from the power
spectra before a best frequency is chosen.  Several well-known
interference signals were automatically removed from all FFTs (most
notably the 19.2 Hz VLA data valid signal and 60 Hz commercial power,
and their harmonics).  Other interference was found by analyzing
groups of sources simultaneously and comparing the best frequencies
found by the search code.  Matching frequencies found in two or more
independent pointings were successively eliminated before folding the
time series.

\section{Sensitivity and Results}

Our two search methods have different sensitivities.  The continuum
maps are limited by the thermal noise of the VLA electronics.  The
HTRP's sensitivity is additionally limited by dispersive smearing of
the pulse across the finite bandpass of the VLA and by scattering.
However, in the absence of significant dispersion and scattering, the
short duty cycles of most pulsars potentially allow pulse searches to
have higher sensitivity than the VLA imaging counterpart.

With 12 minutes of data for each source at each frequency, the thermal
noise limits for the VLA at 1.4 and 5.0 GHz are roughly 0.13 mJy and
0.10 mJy, respectively (Bridle 1989).  In practice, our noise level
was higher because of the unusual hybrid array configuration.  Our
average noise level at 1.4 GHz was $\simeq 0.2$ mJy, giving a
3-$\sigma$ detection limit of 0.6 mJy.

To calculate our sensitivity to a pulsed signal (as a function of
pulse period), we used Eq.\ (9) from Nice et al.\ (1995) with the
appropriate values of average system noise, observing frequency, and
filter bandwidth for the VLA. This ``sensitivity curve'' (Figure 1) is
a plot of minimum detectable flux density (in mJy) versus pulse period
(in ms).  The sensitivity is also a function of DM.  We used the
galactic electron distribution model (and code) of Taylor and Cordes
(1993) to calculate a DM for each observed OB runaway (see Table 1).
Assuming a time domain SNR of 5 for an integrated pulse profile, a
small duty cycle ($< 2\%$) and a maximum of 32 harmonics, we calculate
an average sensitivity of 0.2 mJy for long period pulsars (P $\ge 50$
ms).

To confirm the above calculation, we introduced simulated pulsars with
Gaussian profiles of random amplitudes and periods into one of our
time series and then searched the data for pulses, without a priori
knowledge of the amplitude, period, or even presence of a simulated
pulsar.  The sensitivity curve predicted well our ability to discover
simulated pulsars.  As a final check, two of the known pulsars which
we observed, PSR J1804-0735 (P = 23 ms) and PSR J0017+5914 (P = 101
ms), are sub-mJy sources (0.5 and 0.3 mJy respectively).  Both were
detected with high confidence.

Out of 44 program stars, we detected no pulsars and no non-thermal
point sources.  If we assume that pulsar companions to OB runaways
have properties similar to young, non-recycled pulsars, then our
search would have been sensitive to the vast majority of any such
pulsars if they are not obscured by a massive wind from the OB star.

\section{Model Assumption and Statistics}

A search of a given OB runaway may or may not turn up a pulsar.  We
denote by $O_p$ the event of observing a pulsar and by
$\overline{O}_p$ the failure to detect a pulsar.  Independently of the
issue of detection, we denote by $E_p$ the event of there being a
potentially detectable pulsar associated with the OB star, where
pulsar is here defined to mean the existence of a radio pulse emitting
neutron star with its beam directed toward us that is sampled from a
set of pulsars with some luminosity distribution $\cal{L}$.
Conditional probabilities exist for observing a pulsar or not subject
to the existence of a pulsar.  In principle, the probability
$P(O_p|\overline{E}_p)$ might be nonzero if an unrelated field pulsar
happens to lie within the VLA's beam or one of its sidelobes.  In
practice, the VLA beam is small and the chance of this occuring in any
one pointing is small.  In the analysis to follow we ignore
unassociated field pulsars and set $P(O_p|\overline{E}_p)=0$.
Depending upon the assumed luminosity distribution $\cal{L}$ of OB
runaway pulsar companions, the distance to the star and the flux
sensitivity, there will be some probability
$P_k(\overline{O}_p|E_p,{\cal L},I_k)$ of {\it not} observing a pulsar
even if one exists in any pointing $k$.  Here $I_k$ encapsulates
instrumental and source properties, such as flux sensitivity and
distance, that contribute to detectability.

Not every OB runaway will have a pulsar companion and we denote by
$f_p$ the fraction that do.  There will then be some probability of
obtaining a given observation $\theta_k$ from a set of observations
$\{\theta\}$ given a value of $f_p$.  For example, the probability of
not observing a pulsar ($\theta_k=\overline{O}_p$) in pointing $k$ is
\begin{equation}
P_k(\overline{O}_p|f_p,{\cal L},I_k) = f_p
P_k(\overline{O}_p|E_p,{\cal L},I_k) + 1 - f_p
\end{equation}
since we may miss a weak pulsar, or there may be no pulsar to observe.

Bayes' formula can be used to obtain a probability distribution for $f_p$:
\begin{equation}
P(f_p|\{\theta\},{\cal L},\{I\}) = {P(\{\theta\}|f_p,{\cal L},\{I\})
\over P(\{\theta\}|{\cal L},\{I\})} P(f_p) ,
\end{equation}
where $P(f_p|\{\theta\},{\cal L},\{I\})$ is the probability density of
$f_p$ given a set of observations $\{\theta\}$, our model (assumed)
luminosity distribution, ${\cal L}$, and instrumental and source
factors, $\{I\}$.  The probability of the set of observations, given
$f_p$, is $P(\{\theta\}|f_p,{\cal L},\{I\})$ and $P(f_p)$ denotes any
prior knowledge of $f_p$.  The probability density
$P(f_p|\{\theta\},{\cal L},\{I\})$ for $f_p$ is normalized by the
factor $P(\{\theta\}|{\cal L},\{I\})$.

We failed to detect a pulsar in every pointing, so $\theta_k =
\overline{O}_p$ for all $k$.  Since each of our observations yielded a
null result, we need to compute $P_k(\overline{O}_p|E_p,{\cal L},I_k)$
using an assumed luminosity distribution and our sensitivity.  As a
model we assume that pulsar companions to OB runaways would mirror the
properties of known, young, non-recycled pulsars.  Using the recorded
1400 MHz luminosities of these pulsars and our sensitivity curve we
compute the number of pulsars $N_{\overline{O}_p}(I_k)$ that would
have gone undetected in our search.  With this we can estimate the
probability by
\begin{equation}
P_k(\overline{O}_p|E_p,{\cal L},I_k) = {N_{\overline{O}_p}(I_k) \over
N_{1400}} ,
\end{equation}
where $N_{1400} = 353$ is the total number of young, non-recycled
pulsars with recorded flux at 1400 MHz.  Figure 2 presents a sample
calculation.  Computed values of $P_k(\overline{O}_p|E_p,{\cal
L},I_k)$ are listed in Table 1.  Note that, for most of our program
stars, $P_k(\overline{O}_p|E_p,{\cal L},I_k) = 0$, reflecting the high
sensitivity of our observations.

Since each of our observations was independent, the total probability
$P(\{\theta\}|f_p,{\cal L},\{I\})$ is the product
\begin{equation}
P(\{\theta\}|f_p,{\cal L},\{I\}) = \prod_{k=1}^{N}
P_k(\overline{O}_p|f_p,{\cal L},I_k) .
\end{equation}
Assuming no prior knowledge of $f_p$ and normalizing so that $\int
P(f_p|\{\theta\},{\cal L},\{I\}) df_p = 1$, we find that our null
result bounds $f_p$ to be
\begin{equation}
f_p \le 0.0651 ,
\end{equation}
with 95\% confidence.  Figure 3 shows a plot of
$P(f_p|\{\theta\},{\cal L},\{I\})$ versus $f_p$.  To test our result's
sensitivity to uncertainties in the low-luminosity tail of the pulsar
luminosity distribution, we repeated the calculation assuming twice
and half our stated sensitivity.  The 95\% confidence interval for
$f_p$ varied between 0.0647 and 0.0661, respectively, reflecting the
fact that our survey was sensitive enough to detect the vast majority
of pulsars consistent with our model luminosity function.  Our result
is consistent with the findings of Sayer et al.\ (1995) who conducted
a similar search.

In order to draw conclusions about the mechanism for producing OB
runaways, we must distinguish between pulsars and neutron stars.  Four
main factors could cause a NS not to be classified as a pulsar: the
pulsar may never have turned on, the pulses may be obscured, they may
be beamed away from us, they may have ceased because of old age.  If
we denote the fraction of neutron stars which never turn on by $f_x$,
the fraction of pulsar companions to OB runaways which are obscured by
$f_o$, the fraction that have fallen below the death line by $f_d$,
and the fraction which are beamed toward us by $f_b$, then
\begin{equation}
f_p = f_{NS} f_b (1 - f_x - f_o - f_d) ,
\end{equation}
where $f_{NS}$ is the fraction of OB runaways having NS companions.
It is difficult to set limits on $f_o$.  However, PSR B1259-63 shows
dispersive delays only very close to periastron passage (Johnston et
al.\ 1995), indicating that the relativistic pulsar wind comes into
pressure balance with the stellar wind at a distance from the NS
surface far enough to leave the pulsar undisturbed for most of the
orbit.  If the majority of pulsars in such systems have large
spin-down luminosities, then $f_o$ should be small.  OB lifetimes
range from a few to ten million years, comparable to the lifetime of a
single, non-recycled pulsar.  However, the lifetime of a supernova
ejected OB runaway will likely be much shorter due to the
pre-supernova stage of binary stellar evolution.  Consequently, one
can anticipate $f_d \ll 1$.  The beaming fraction $f_b$ is usually
taken to be of order \onethird\ (Lyne \& Graham-Smith 1990).  Setting
$f_x = f_o = f_d = 0$, our observations constrain $f_{NS} < 0.2$.

\section{Discussion: Implications for Runaway Star Formation and Natal
Kicks of Neutron Stars}

The lower than expected observed fraction of pulsar companions to the
OB runaway stars has at least three interpretations.  Firstly,
supernova ejection may be the dominant process of ejecting stars from
OB associations and those stars that retain NS companions are the
HMXBs, combined with the Be and WR stars with NS companions, while the
classical OB runaways represent merely the unbound minority of
supernova-ejected systems.  Secondly, supernova ejection may again be
the dominant process producing runaways but the natal kicks that NSs
receive at birth are larger than has been previously assumed which
results in mostly single runaways.  Thirdly, the majority of the OB
runaways are dynamically ejected rather than supernova ejected.

It has been proposed that HMXBs are runaway stars (van Oijen 1989;
Brandt \& Podsiadlowski 1995), and so perhaps these combined with the
Be stars with NS companions (Johnston et al.\ 1992; Kaspi et al.\
1994; Schmidtke et al.\ 1995) and runaway WR stars with compact
companions (Moffat, Lamontagne \& Seggewiss 1982; Isserstedt, Moffat
\& Niemela 1983; Robert et al.\ 1992) provides a group large enough to
account for the long sought after supernova-ejected runaways with NS
companions.  In this scenario, the classical OB runaways would
represent the minority of runaway systems, from which the NSs have
escaped.  However, this is unlikely, since the classical OB runaways
appear to outnumber the other three types of stars put together.

The observed lack of NS companions to the OB runaways may be due to
larger than expected kicks given to NSs at birth.  Indeed, the
evidence for large natal kicks is growing (Frail \& Kulkarni 1991;
Cordes, Romani \& Lundgren 1993; Lyne \& Lorimer 1994; Frail, Goss \&
Whiteoak 1994).  The pioneering Monte Carlo simulations of Dewey \&
Cordes (1987) suggest that a mean three-dimensional kick of 90 km
s$^{-1}$ is required to reproduce the observed velocity distribution
and binary properties of the then-known pulsars.  Curve~E in Figure~2
of the more recent study of Leonard, Hills \& Dewey (1994) shows that
a one-dimensional kick dispersion of 240 km s$^{-1}$ (a mean
three-dimensional kick of 420 km s$^{-1}$) is required to disrupt
$\simeq 80$\% of runaway OB plus NS systems.  This number is similar
to the recent estimate of the mean pulsar velocity of $450 \pm 90$ km
s$^{-1}$ found in the study of Lyne \& Lorimer (1994).  Thus, our
result of 20\% for $f_{NS}$ is in very close agreement with recent
revisions of the pulsar velocity distribution.

Alternatively, dynamical ejection could account for the majority of
the OB runaway stars.  It has been argued that this mechanism
naturally accounts for many of the observed properties of the OB
runaways (Gies \& Bolton 1986; Gies 1987), including that the fastest
dynamically-ejected runaways are expected to be single stars (Leonard
\& Duncan 1988, 1990; Leonard 1995).  The fastest supernova-ejected
systems may, in fact, be the HMXBs and perhaps the WRs with compact
companions.  The number of ``normal looking'' supernova-ejected
runaway stars could be greatly outnumbered by those that result from
dynamical ejection.

\acknowledgments

C.J.P.~would like to thank NRAO-VLA for hospitality in summer 1993 and
David Nice for helpful discussion about sensitivity.  This research
was supported by NSF grants PHY 90-57865 and PHY 93-17638, and by NASA
grant NAGW 2934.  C.R.E.~thanks the Alfred P.~Sloan Foundation for
research support.  P.J.T.L.~is grateful to the University of Maryland
for financial support.  This research made use of the Simbad database,
operated at CDS, Strasbourg, France.

\clearpage

\begin{deluxetable}{lllllll}
\tablecaption{Observed OB Runaways}
\tablehead{
\colhead{OB Runaway}           & \colhead{Distance}      &
\colhead{$V_p$} & \colhead{Type}  &
\colhead{DM$_{\rm calc}$}      &\colhead{$P_k$} &
\colhead{ref}}
\startdata
72 Col & 0.26 & 191 & B2.5V & 4.8 & 0.000 &1 \\
AGK+60$^o$1562 & 0.55-0.7 & 40 & B9III  & 13.1 & 0.000  &1 \\
BD+59 186 & 1.8-3.1 & 56 & B5II  & 83.5 & 0.067  &1 \\
HD 4142 & 0.26 & -60 & B4V  & 4.9 &  0.000  &4 \\
HD 14220 & 0.61 & -34 & B2V  & 11.1 &  0.000  &4 \\
HD 16429 & 1.39 & -24 & O9.5III  & 25.6 & 0.008   &4 \\
HD 17114 & 0.7-0.9 & 63 & B1V &  16.7 & 0.000 & 1 \\
HD 19374 & 0.36 & 59 & B1.5V  & 6.6 & 0.000  &1 \\
HD 20218 & 1.0--1.4 & 40 & B2V  & 25.5 & 0.008 &1 \\
HD 24912 & 0.51 & 58 & O7.5 & 9.4 & 0.000 &4 \\
HD 29866 & 0.29 & -3 & B8IVn  & 5.4 & 0.000  &4 \\
HD 30614 & 1.25 & 26 & O9.5Ia  & 22.1 & 0.000  &1,2 \\
HD 30650 & 0.41 & 34 & B6V  & 7.6 & 0.000  &4 \\
HD 34078 & 0.52 & 49 & O9.5V  & 9.7 & 0.000  &2,4 \\
HD 37737 & 2.16 & 25 & O9.5III  & 56.0 & 0.025   &4 \\
HD 38666 & 0.7 & 123 & O9.5V  & 12.5 & 0.000  &1,2 \\
HD 39478 & 0.85--1.2 & 49 & B2V  & 22.0 &  0.000 &1 \\
HD 39680 & 2.48 & 29 & O6V &  55.5 & 0.039 &4 \\
HD 43112 & 0.65 & 19 & B1V  & 12.0 & 0.000  &4 \\
HD 52533 & 1.82 & 9.6 & O9V  & 33.1 &  0.014  &4 \\
HD 78584 & 0.89 & 111 & B3 & 15.0 & 0.000 &1 \\
HD 91316 & 0.59 & 36 & B1Ib  &  8.3 & 0.000 &3 \\
HD 97991 & 0.93 & 22 & B2V  & 14.3 &  0.000 &4 \\
HD 125924 & 2.47 & 259 & B2IV  & 20.6 & 0.014 & 3 \\
HD 149363 & 1.11 & 148 & B0.5V & 19.2  & 0.000 &4 \\
HD 149757 & 0.17 & 39 & 09.5  & 3.2 &  0.000 &1,2 \\
HD 157857 & 2.40 & 50 & O7f & 54.0 &  0.039 &1 \\
HD 172488 & 0.54 & 41 & B0.5v  & 10.3 &  0.000 &4  \\
HD 188439 & 1.08 & -58 & B0.5IIIp & 20.0 & 0.000  &4 \\
HD 189957 & 2.55 & 49 & O9.5III & 46.7 & 0.039 &4 \\
HD 192281 & 1.78 & -32 & O5e  & 33.5 & 0.014 &4 \\
HD 195907 & 0.92 & -68 & B1.5Ve  & 17.2 & 0.000 &4 \\
HD 201345 & 1.92 & 29 & ON9V  & 34.7 &  0.014  &4 \\
HD 201910 & 0.54 & 4 & B5V  & 10.1 &  0.000  &4 \\
HD 203064 & 0.85 & 20 & O7.5 & 16.0 & 0.000 &4 \\
\tablebreak
HD 210839 & 0.86 & -54 & O6Iab  & 16.1 & 0.000  &1,2 \\
HD 214080 & 2.40 & 32 & B1/B2Ib  & 19.0 & 0.014 &3 \\
HD 214930 & 0.75 & -46 & B2IV  & 13.3 & 0.000  &4 \\
HD 216534 & 0.83 & 82 & B3V  & 15.4 & 0.000  &1 \\
HD 219188 & 2.11 & 82 & B0.5III  & 19.1 &  0.011  &3 \\
HD 220172 & 0.77 & 18 & B3Vn  & 11.1   &  0.000  &3 \\
HD 248434 & 3.0-6.0 & 83 & B5Ie & 113.1 & 0.204  &1 \\
HD 276968 & 1.3-2.0 & 55 & B9II  & 54.4 & 0.014 &1 \\
HD 333282 & 1.3  & 40 & B7III  & 24.7 & 0.003 &1 \\
\tablerefs{
(1) Bekenstein \& Bowers 1974; (2) Blaauw 1993; (3) Conlon et al.\ 1990;
(4) Gies \& Bolton 1986.}
\tablecomments{Distances are in kpc.  $V_p$ are in km s$^{-1}$.
DM values, in pc cm$^{-3}$, were calculated using the code
of Taylor and Cordes (1993).  Where a range of distances is given, the
larger was used to calculate the DM.  $P_k$ refers to the probability,
$P_k(\overline{O}_p|E_p,{\cal L},I_k)$, of not observing a pulsar subject
to the existence of a pulsar, the assumed luminocity distribution ${\cal L}$,
and instrumental factors $I_k$.}
\enddata
\end{deluxetable}

\begin{deluxetable}{lllllll}
\tablecaption{Observed Known Pulsars}
\tablehead{
\colhead{Pulsar}           & \colhead{Period}      &
\colhead{S$_{1400}$} & \colhead{DM}}
\startdata
J0117+5914 & 101.44 & 0.3 & 48.5 \\
J0147+5922 & 196.32 & 2.0 & 39.3 \\
J1804-0735 & 23.10 & 0.5 & 186.4 \\
J1833-0827 & 85.28 & 5.0 & 411.0 \\
\tablecomments{Pulse periods are in milliseconds.  S$_{1400}$ is the time
average flux density, in mJy, at 1400 MHz.  The measured DM values are in
pc cm$^{-3}$. All values are from Taylor et al.\ (1993)}
\enddata
\end{deluxetable}

\begin{figure}
\plotone{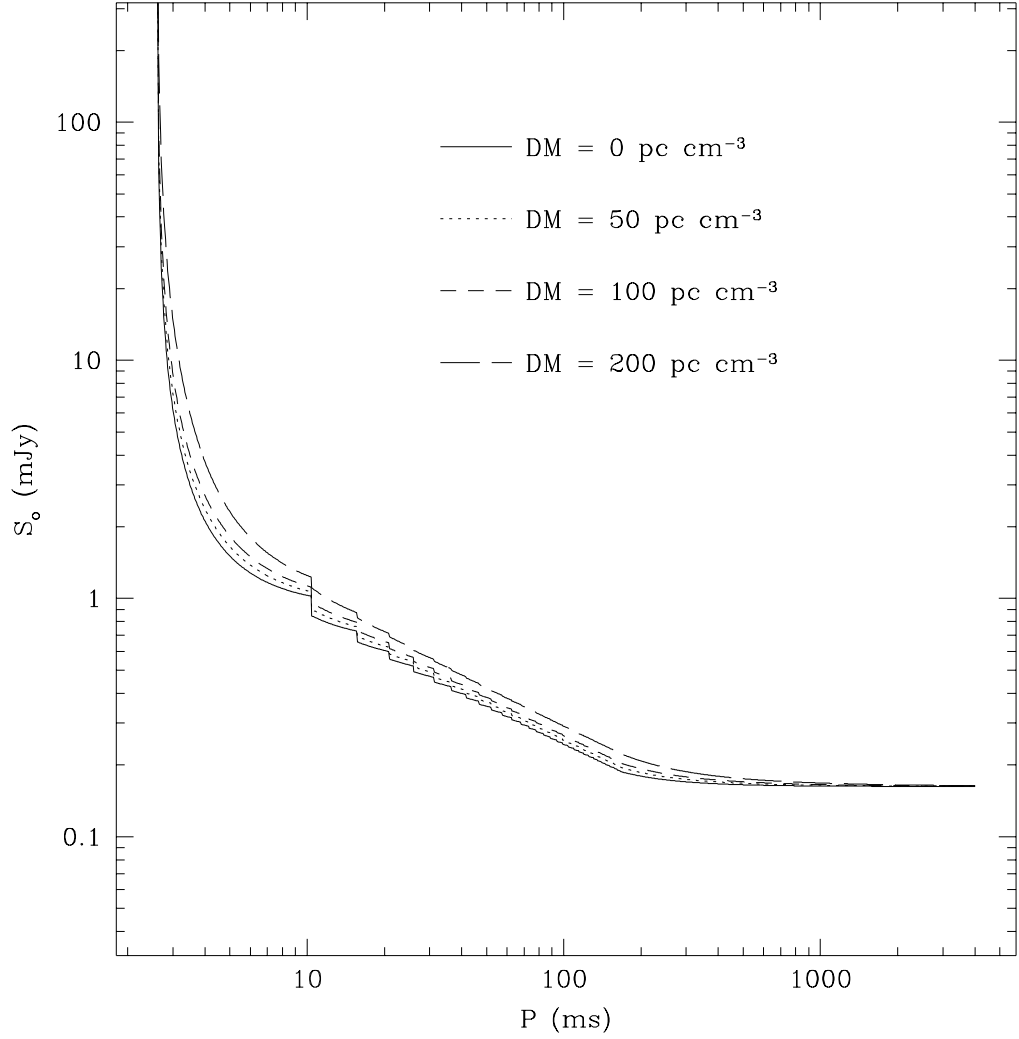}
\caption{Pulsar Sensitivity.  The lowest detectable flux density, S$_o$, is
plotted as a function of pulse period.  These curves were calculated using
Eq.\ (9) from Nice et al.\ 1995.  We assumed a signal-to-noise ratio of five
(in the time domain), a small duty cycle ($\le$ 2\%) and a maximum of 32
harmonics in the power spectrum.}
\end{figure}

\begin{figure}
\plotone{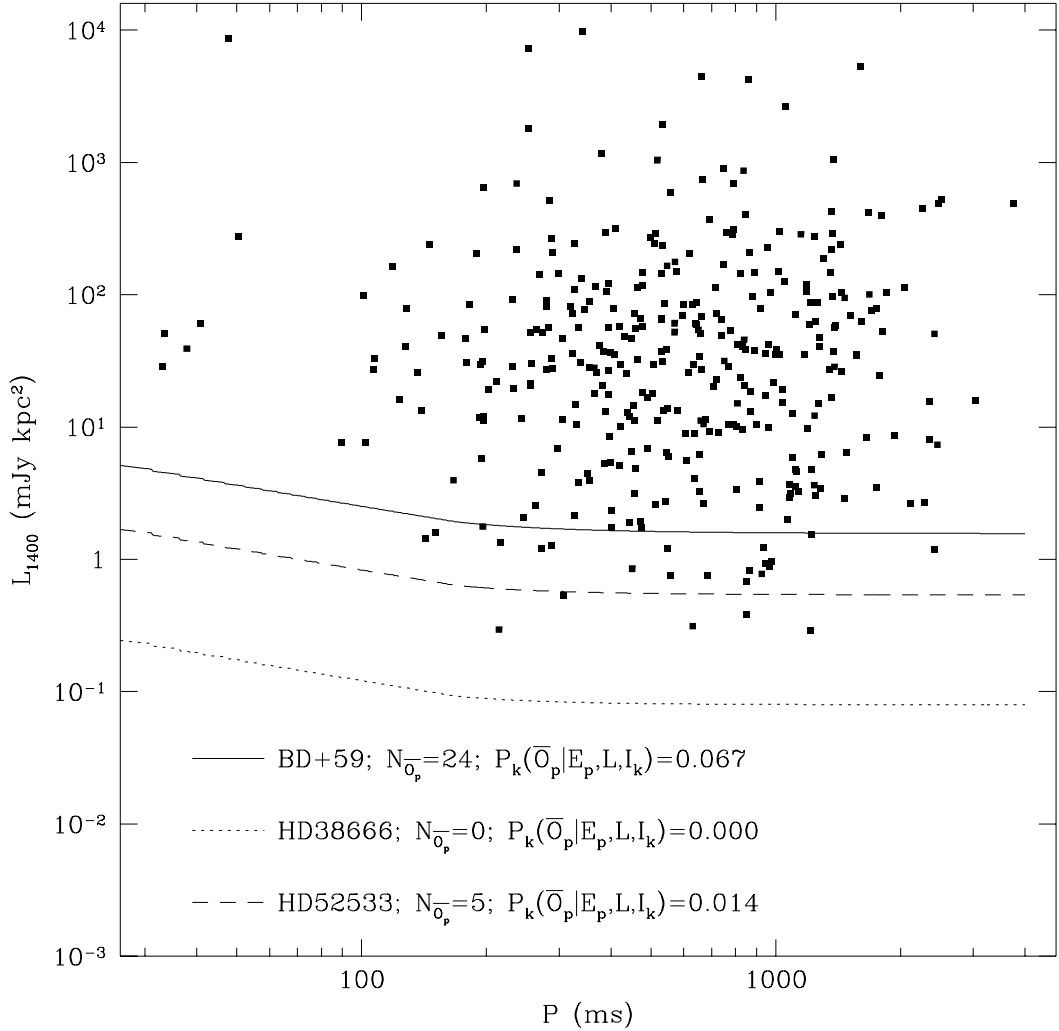}
\caption{Calculation of conditional probability.  The squares mark the
1400 MHz luminosities (flux density times distance squared) of the known
pulsars (Taylor et al.\ 1993).  The recycled or millisecond pulsars have
been excluded because they represent a separate population.  Here, the
sensitivity curves (calculated for the DMs in Table 1) are scaled by the
distance in order to reflect luminosity sensitivity.  $N_{\overline{O_p}}$
represents the number of known pulsars that fall below a given sensitivity
curve.}
\end{figure}

\begin{figure}
\plotone{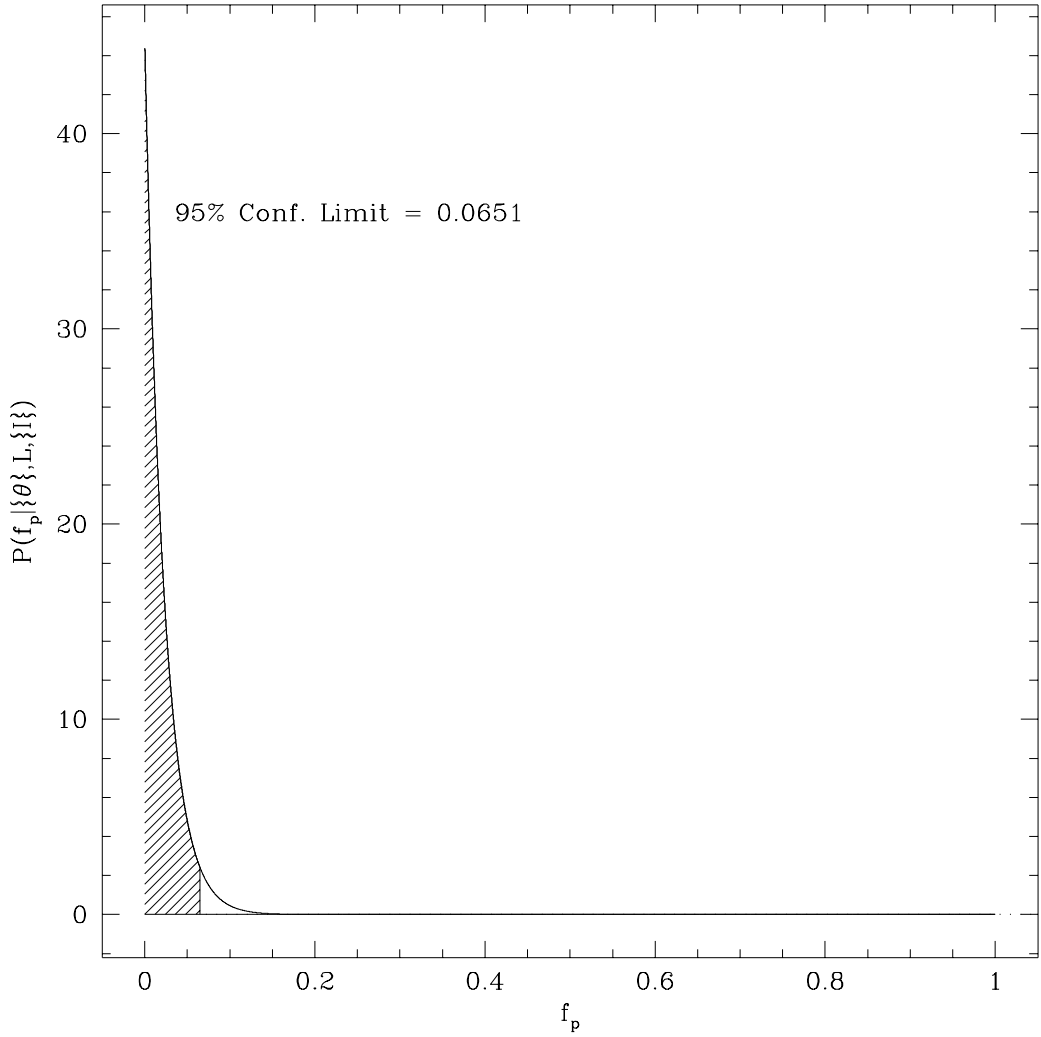}
\caption{Probability Density of $f_p$.
$P(f_p|\{\theta\},{\cal L},\{I\})$ is normalized so that the total area under
the curve equals 1.0.  The 95\% confidence limit is the value of $f_p$ to the
left of which the area under the curve equals 0.95 (the shaded area).}
\end{figure}

\end{document}